\documentclass[journal]{IEEEtran}

\usepackage{amssymb,amsmath}
\usepackage{url}

\usepackage{graphicx}
\usepackage{cite}

\newcommand{\be}{\begin{equation}}
\newcommand{\ee}{\end{equation}}
\newcommand{\Dlt}{\Delta}
\newcommand{\dlt}{\delta}

\newcommand{\bt}{\beta}
\newcommand{\vp}{\varphi}

\newcommand{\al}{\alpha}
\newcommand{\ra}{\rightarrow}

\newcommand{\Gm}{\Gamma}

\newcommand{\cH}{{\cal H}}
\newcommand{\cL}{{\cal L}}

\newcommand{\rgl}{\rangle}
\newcommand{\lgl}{\langle}

\begin{document}

\title{Manipulating decision making of typical agents}

\author{Vyacheslav I.~Yukalov
        and~Didier~Sornette 

\thanks{V.I. Yukalov and D. Sornette are with the Department of Management,
Technology and Economics, ETH Z\"urich, Swiss Federal Institute of Technology,
Z\"urich CH-8032, Switzerland.}

\thanks{V.I. Yukalov is with the Bogolubov Laboratory of Theoretical Physics,
Joint Institute for Nuclear Research, Dubna 141980, Russia (e-mail:
yukalov@theor.jinr.ru)}

\thanks{D. Sornette is with the Swiss Finance Institute, c/o University of Geneva,
40 blvd. Du Pont d'Arve, CH 1211 Geneva 4, Switzerland
(e-mail: dsornette@ethz.ch)}

\thanks{Manuscript received ; revised }}

\markboth{IEEE Transactions on Systems Man $\&$ Cybernetics: Systems,~Vol.~, No.~, ~}%
{Shell \MakeLowercase{\textit{et al.}}: Bare Demo of IEEEtran.cls for Journals}

\maketitle

\begin{abstract}
We investigate how the choice of decision makers can be varied under the presence
of risk and uncertainty. Our analysis is based on the approach we have previously
applied to individual decision makers, which we now generalize to the case of
decision makers that are members of a society. The approach employs the
mathematical techniques that are common in quantum theory, justifying our naming
as Quantum Decision Theory. However, we do not assume that decision makers
are quantum objects. The techniques of quantum theory are needed only for defining
the prospect probabilities taking into account such hidden variables as behavioral
biases and other subconscious feelings. The approach describes an agent's
choice as a probabilistic event occurring with a probability that is the sum of
a utility factor and of an attraction factor. The attraction factor embodies
subjective and unconscious dimensions in the mind of the decision maker. We show
that the typical aggregate amplitude of the attraction factor is $1/4$, and it can
be either positive or negative depending on the relative attraction of the
competing choices. The most efficient way of varying the decision makers choice is
realized by influencing the attraction factor. This can be done in two ways. One
method is to arrange in a special manner the payoff weights, which induces the
required changes of the values of attraction factors. We show that a slight
variation of the payoff weights can invert the sign of the attraction factors and
reverse the decision preferences, even when the prospect utilities remain unchanged.
The second method of influencing the decision makers choice is by providing
information to decision makers. The methods of influencing decision making are
illustrated by several experiments, whose outcomes are compared quantitatively
with the predictions of our approach.
\end{abstract}

\begin{IEEEkeywords}
Decision theory, Decision making under risk and uncertainty,
Group consultations, Social interactions, Information and knowledge
\end{IEEEkeywords}

%
\IEEEpeerreviewmaketitle

\section{Introduction}

How to influence decision choices made by separate decision makers as well
as by societies of many agents is an important and widely studied problem
in psychology \cite{1,2,3,4,5,6}. This problem is important for a variety
of practical applications ranging from medicine \cite{7,8} to politics
\cite{9,10,11,12,13}. A number of articles are devoted to the effects of
influencing decision making in economics, studying the role of different
framing effects on product evaluation \cite{14,15,16}, consumer response
to price \cite{17,18,19,20}, evaluation of retail outlets \cite{21},
market advertising \cite{22,23,24}, buying decisions \cite{25,26},
perceptions of control and efficacy \cite{27}, distributive justice
\cite{28}, performance feedbacks \cite{29}, and so on.

The principal possibility of influencing the choices of decision makers is
based on the fact that decision makers do not exactly follow the prescriptions
of expected utility theory, as formulated by von Neumann and Morgenstern
\cite{Neumann_30}. Really, if each decision maker were to make decisions
following the strict rules of utility theory, then it would be difficult,
if possible at all, to influence his/her decisions without essentially
varying the utility of the related lottery. However, it is well known that
the choices of decision makers are not based solely on utility, but also
are strongly influenced by emotions, prejudices, biases, and other
subconscious feelings. There have been numerous attempts to modify utility
theory by taking into account such subconscious degrees of freedom. For this
purpose, a number of non-additive nonlinear probability models have been
developed to account for the deviations from objective to subjective
probabilities observed in human agents
\cite{42,43,Quiggin_30,Gilboa_30,Schmeidler_30,Gilboa_31,Cohen_31,
Montesano_31,40,Shefrin_31}, trying to take into account satisfaction
\cite{Guo_31}, anxiety \cite{Epstein_31}, subjective perception
\cite{West_31}, subjective utility \cite{Blavatskyy_31}, aspiration adaptation
\cite{Selten_31,Napel_31}, and so on. The necessity of taking into account
the subconscious behavioral biases is emphasized in behavioral economics
\cite{Camerer_41}. The variety of the approaches, deviating from the expected
utility theory, are commonly named "non-expected utility theories" \cite{Machina_42}.

The non-expected utility theories have been thoroughly analyzed in several
reviews \cite{Machina_42,Safra_42,Al-Najjar_42,Al-Najjar_43}. The conclusion
is that such theories are in the best case only descriptive, hence, do not
have predictive power and do not explain numerous paradoxes existing in
classical decision making. Moreover, their use ends up creating more paradoxes
and inconsistencies than it resolves \cite{Al-Najjar_42}.

To overcome the problem, we have developed an approach based on the
mathematical techniques of quantum theory \cite{30,31,32,33,34,35,36},
which explains our choice of its name, Quantum Decision Theory (QDT). We do not
assume that decision makers are quantum objects. But the mathematical quantum
techniques serve as the most convenient tool for taking into account the
subconscious degrees of freedom of decision makers, similarly to how quantum
theory avoids the explicit use of hidden variables, at the same time taking
into account their possible existence resulting in the probabilistic formulation
of the theory. We have shown that, in the frame of QDT, all paradoxes of
classical decision making find simple and natural explanations \cite{32,35,36}.
QDT provides the expressions for discount functions, employed in the
theory of time discounting \cite{Frederick_36,Rambaud_36} and explains dynamical
inconsistences \cite{31}. Within QDT, behavioral biases result from
interference and entanglement caused by decision makers deliberations \cite{34}.
While QDT has been developed to describe the behavior of human decision
makers, it can also be used as a guide for creating artificial quantum
intelligence \cite{33}.

In our previous publications \cite{30,31,32,33,34,35,36}, we have considered a
single decision maker. In the present paper, we generalize QDT, developed
earlier for individual decision makers, to the case of decision makers interacting
within a society. This generalization is formulated in Sec. II. Our main concern
is the formulation of a {\it mathematical model} describing how decision makers
can be influenced and how it would be possible to {\it quantitatively} evaluate
the consequences of this influence. In experiments, one usually deals with large
groups of decision makers with different preferences. In order to compare
theoretical predictions with experimental results, we introduce and characterize,
in Sec. III, the notion of typical social agents. Here and in what follows, by the
term `experiments', we mean empirical observations derived from the behavior of
human subjects. In Sec. IV, we explain how it is possible to influence the typical
decision makers preference by varying the arrangement of prospects. We formulate
a criterion for the inversion of the attraction factor leading to the inversion
of preferences. In Sec. V, the results, predicted by our approach, are compared
with several classical experiments, demonstrating good quantitative agreement.
In Sec. VI, we show how decision makers can be influenced by providing them
additional either correct or wrong information. Section VII concludes.

\section{Quantum decision making in society}

In this section, we generalize the QDT approach, whose detailed exposition
can be found in our previous publications, developed for individual decision
makers, to a society of many decision makers. We recall that the decision makers
are not quantum objects, but are normal humans. The techniques of quantum theory
are employed merely for taking into account the hidden variables, such as emotions
and biases of decision makers \cite{36}. The possibility of taking into account
hidden variables is at the heart of the quantum-theory techniques that allow for
their existence by modifying the rules of calculating the quantum probabilities.
This is why the quantum techniques make it possible to characterize human
decision making, incorporating in it the existence of such hidden variables as
subconscious feelings and behavioral biases. The efficiency of quantum techniques
for human decision making is not because humans are quantum objects, but because
these techniques are mathematically designed to accommodate the existence of
hidden variables, which can be of a very different nature.

The theory presented below requires the use of some mathematical techniques
that are common in quantum theory. But, as is explained above, the reader does
not need to know anything about quantum theory. Actually, what one needs is the
basic knowledge of the functional analysis in Hilbert space and the definition
of scalar products used for introducing the prospect probabilities. In order
to make the presentation self-consistent and to justify the derived results,
we describe the main mathematical steps of the derivation, at the same time,
omitting intermediate calculations for not overloading the reader. However,
we cannot omit all mathematical formulas, since then it would not be clear how
we get the final important expressions and why they have the properties that
are essentially used in the following applications.

Let us consider a society of $N$ agents who are decision makers. The agents
are enumerated by $\alpha = 1,2,\ldots,N$. Each agent is characterized by
a set $\{e_{\alpha n}: n = 1,2,\ldots,d \}$ of $d$ elementary prospects
that are represented by vectors $|\alpha n \rangle$ in a Hilbert space.

Everywhere below, we employ the Dirac \cite{Dirac,38} bracket notation, where
a function $\psi_n(x)$ is represented as $|n\rangle$ and the scalar product
of two functions is given by the formula
$$
\lgl m \; | \; n \rgl \equiv \int \psi_m^*(x) \psi_n(x) \; dx \; .
$$

Different elementary prospects are orthonormal to each other,
$$
\lgl \al m | \al n \rgl = \dlt_{mn} \;   ,
$$
which symbolizes their mutual independence and incompatibility. The space
of mind of an $\alpha$ - decision maker is the Hilbert space
\be
\label{1}
 \cH_\al \equiv {\rm Span}_n \{ | \al n \rgl \} \; .
\ee
The dimension of this space of mind is $d$. The space of mind of the whole
society is the tensor product
\be
\label{2}
 \cH = \bigotimes_{\al=1}^N \cH_\al \;  ,
\ee
whose dimension is $Nd$.

An $\alpha$ - agent deliberates on deciding between $L$ prospects forming
a complete lattice
\be
\label{3}
 \cL_\al \equiv \{ \pi_{\al j} : \; j = 1,2,\ldots, L \} \; .
\ee
Each prospect $\pi_{\alpha j}$ is represented by a vector
$|\pi_{\alpha j} \rangle$ in the space of mind (1). The prospect vectors
do not need to be orthonormal, which implies that they are not necessarily
incompatible.

The prospect operator
\be
\label{4}
\hat P(\pi_{\al j} ) \equiv | \pi_{\al j} \rgl \lgl \pi_{\al j} |
\ee
acts on the space of mind (1). The set of all these operators is analogous
to the algebra of local observables in quantum theory \cite{38}.
Respectively, the prospect probabilities are defined as the expectation
values of the prospect operators. The expectation values for an individual
decision maker are given by averaging the prospect operators over a
strategic state of this decision maker \cite{36}, with such a strategic state 
being treated as a pure state represented by a single vector.

However, for the agents of a society, pure states of separate agents, generally, 
do not exist, since the society agents interact with each other by exchanging
information. Moreover, the society as a whole may not be completely isolated
from its surrounding. Therefore, the society state has to be characterized by
a statistical operator $\hat{\rho}$ that is a non-negative normalized operator,
\be
\label{5}
{\rm Tr}_\cH \hat\rho = 1 \; ,
\ee
where the trace operation is over the society space (2). Then the
expectation values of the prospect operators are given by the trace
\be
\label{6}
 p(\pi_{\al j}) \equiv {\rm Tr}_\cH \hat\rho \hat P(\pi_{\al j} ) \; ,
\ee
defining the probabilities of the corresponding prospects. This
definition makes the basic difference in the calculation of the prospect
probabilities, as compared to the averaging over a single strategic state
for individual decision makers \cite{30,31,32}.

Quantity (6), by its construction, is non-negative and defines the
prospect probabilities under the normalization condition
\be
\label{7}
 \sum_{j=1}^L p(\pi_{\al j} ) = 1 \; , \qquad
0 \leq p(\pi_{\al j} ) \leq 1 \;  .
\ee
This definition of prospect probabilities is similar to the definition
of quantum probabilities in the quantum theory of measurements \cite{YS}.

Remembering that the prospect operator (4) acts on the space of
mind (1) and introducing the reduced statistical operator
$$
\hat\rho_\al \equiv {\rm Tr}_{\cH /\cH_\al} \hat\rho \; ,
$$
in which the trace is over the partial factor space
$$
\cH /\cH_\al \equiv \bigotimes_{\bt(\neq\al)}^N \cH_\bt \;  ,
$$
makes it possible to rewrite the prospect probability (6) in the form
\be
\label{8}
p( \pi_{\al j}) = {\rm Tr}_{\cH_\al}\hat\rho_\al
\hat P(\pi_{\al j} ) \;  ,
\ee
with the trace over the space of mind (1).

Expanding the prospect vectors over the elementary prospect basis,
and introducing the matrix elements
$$
\rho_{mn}^\al \equiv
\lgl \al m | \hat\rho_\al | \al n \rgl \; ,
$$
\be
\label{9}
P_{mn}(\pi_{\al j} ) \equiv
\lgl \al m | \hat P(\pi_{\al j}) | \al n \rgl \;   ,
\ee
it is straightforward to get the prospect probability
\be
\label{10}
 p(\pi_{\al j}) = f(\pi_{\al j} ) + q(\pi_{\al j} ) \;  ,
\ee
consisting of two terms. The first term, called the {\it utility factor},
\be
\label{11}
 f(\pi_{\al j} ) =
\sum_n \; \rho_{nn}^\al P_{nn}(\pi_{\al j} ) \; ,
\ee
describes the classical objective probability, showing how the considered
prospect is useful for the decision maker. While the second term,
called the {\it attraction factor},
\be
\label{12}
 q(\pi_{\al j} ) =
\sum_{m\neq n} \; \rho_{mn}^\al P_{nm}(\pi_{\al j} ) \;  ,
\ee
characterizes the subjective influence of subconscious feelings, emotions,
and biases and shows to what extent the prospect is attractive for the
decision maker.

One could think that the form of probability (10) could be postulated,
without deriving it from the preceding equations. Then, however, one would
not know the properties of the terms $f(\pi)$ and $q(\pi)$. Hence, these
properties should also be postulated, thus making the whole consideration
overloaded by a number of postulates. Using the quantum techniques, we obtain
the form of probability (10) automatically, which makes the approach
self-consistent and well justified. In this way, the appearance of two terms
in probability (10) is not an assumption, but it is the straightforward
consequence of using quantum techniques. The properties of these terms follow
directly from Eqs. (1) to (8).

By its definition, the utility factor (11) is non-negative,
\be
\label{13}
 0 \leq  f(\pi_{\al j} ) \leq 1 \;  ,
\ee
and also it is normalized,
\be
\label{14}
 \sum_{j=1}^L f(\pi_{\al j} ) = 1\; ,
\ee
representing the classical objective probability. In the case when the
prospect utilities $U(\pi_{\alpha j})$ can be evaluated by means of
classical utility theory, the utility factor takes the form
\be
\label{15}
 f(\pi_{\al j} ) = \frac{U(\pi_{\al j}) }{\sum_j U(\pi_{\al j}) }\;  .
\ee

The attraction factor (12), by its definition, varies in the range
\be
\label{16}
 -1 \leq  q(\pi_{\al j} ) \leq 1 \;  .
\ee
An important property of the attraction factor, following from conditions
(7) and (14), is the {\it alternation property}
\be
\label{17}
 \sum_{j=1}^L q(\pi_{\al j} ) = 0\;  .
\ee

It is worth mentioning that the attraction factor comes into play only
for composite prospects experiencing mutual interference \cite{36}. For
elementary prospects, it does not occur, being identically zero:
$$
 q(e_{\al n} ) = 0\;  .
$$

Having defined the prospect probabilities, the prospects become naturally
ordered. A prospect $\pi_{\alpha 1}$ is said to be preferred to a prospect
$\pi_{\alpha 2}$ if and only if
\be
\label{18}
 p(\pi_{\al 1}) >  p(\pi_{\al 2}) \qquad
( \pi_{\al 1} > \pi_{\al 2})\;  .
\ee
The prospects $\pi_{\alpha 1}$ and $\pi_{\alpha 2}$ are indifferent if
and only if
\be
\label{19}
 p(\pi_{\al 1}) =  p(\pi_{\al 2}) \qquad
( \pi_{\al 1} = \pi_{\al 2})\;  .
\ee
And the prospect $\pi_{\alpha 1}$ is preferred or indifferent to
$\pi_{\alpha 2}$ if
\be
\label{20}
 p(\pi_{\al 1}) \geq  p(\pi_{\al 2}) \qquad
( \pi_{\al 1} \geq \pi_{\al 2})\;  .
\ee

A prospect $\pi_\alpha^*$ that corresponds to the maximal probability
$$
p(\pi_\alpha^*) =\max_j p(\pi_{\al j})
$$
is called optimal.

It is important to stress that the utility factor and attraction factor
are principally different, having different mathematical properties, as
described above. The term $f(\pi)$ contains only diagonal elements in
sum (11), while term (12) contains only non-diagonal elements. In
quantum theory, the non-diagonal terms characterize the existence of
interference. When the latter is absent, the quantity $f(\pi)$ reduces
to the classical probability. Similarly, in decision theory the term
$f(\pi)$ is associated with the classical probability, while the second
term $q(\pi)$ has no classical counterparts. The attraction factor in
decision theory describes the interference of different prospect modes,
which is related to the deliberation of a decision maker choosing between
several admissible possibilities.

Being principally different from both mathematical as well as
decision-making points of view, the utility and attraction factors in no
way could be combined into one quantity. Actually, the failure of the
numerous ``non-expected utility theories" \cite{Machina_42} is due to the
fact that in these approaches one has tried to construct a single quantity
generalizing the expected utility, which has been shown to be impossible
\cite{Machina_42,Safra_42,Al-Najjar_42,Al-Najjar_43}. In the frame of the
approach of the present paper, it is clear why such a combination of
$f(\pi)$ and $q(\pi)$ is impossible, since they possess very different
mathematical properties. And this impossibility is also easily understood
in the frame of decision theory, where $f(\pi)$ describes an objective quantity
that can be objectively measured, while $q(\pi)$ represents a subjective
quantity that for a single decision maker can be found only empirically,
though its aggregate value for a typical decision maker can be estimated
as is explained in the following section.

The attraction factor in QDT is also basically different from the visceral
factors considered in decision-making literature \cite{Loewenstein}, where
the visceral factors are assumed to be additional unknown variables entering
the definition of utility functions. However, the explicit dependence of such
utility functions on these visceral factors is also not known. Contrary to
this, the properties of the attraction factor are prescribed by its derivation.
In addition, including the visceral factors into utility functions leads to
a redefinition of expected utility combining objective and subjective
features, which is impossible, as discussed above.

\section{Typical behavior of social agents}

Considering large societies, consisting of many agents $N \gg 1$ and
confronting numerous prospects, it is important to understand the typical
behavior of such complex societies, corresponding to their behavior on
average. The society is treated to be large, when $N$ is greater than $10$.
Strictly speaking, the considered society has to contain so many members,
for which it is admissible to collect reliable and representative statistical
data, with a small standard error. For instance, measuring a quantity whose
mean value is $M$, the typical statistical error for $10$ agents is of the
order of $0.3 M$ and for $100$ agents, of order $0.1 M$.

\subsection{Definition of typical agent behavior}

Let all agents in a society confront the same prospect lattice (3), with
the same prospects $\pi_j = \pi_{\alpha j}$. The agents composing
the society are different individuals and their decisions, even related
to the same set of prospects, can vary, producing different probabilities
$p(\pi_{\alpha j})$.

The society as a whole can be characterized by the average probability
\be
\label{21}
 p(\pi_j) \equiv \frac{1}{N} \sum_{\al=1}^N p(\pi_{\al j})\;  ,
\ee
averaged over all society members, which describes the typical behavior
of agents. In view of expression (10), the typical probability (21)
reads as
\be
\label{22}
   p(\pi_j) =  f(\pi_j) +  q(\pi_j) \; ,
\ee
with the typical utility factor
\be
\label{23}
f(\pi_j) \equiv \frac{1}{N} \sum_{\al=1}^N f(\pi_{\al j})
\ee
and typical attraction factor
\be
\label{24}
 q(\pi_j) \equiv \frac{1}{N} \sum_{\al=1}^N q(\pi_{\al j})\;  .
\ee

Expression (22), with terms (23) and (24), appears here directly from
using Eq. (10). That is, the occurrence of the attraction factor (24)
is not an assumption, but the immediate consequence of the employed
mathematical techniques, which themselves embody the entanglement
of composite prospects. The appearance of such additional terms is
typical of quantum theory, where they describe interference effects.

Because of Eqs. (13) and (14), the typical utility factor, describing
the objective probability, satisfies the conditions
\be
\label{25}
 \sum_{j=1}^L f(\pi_j) = 1 \; , \qquad  0 \leq f(\pi_j) \leq 1 \;.
\ee
In the case when it is defined by the prospect utilities according to
Eq. (15), it reduces to the expression
\be
\label{26}
f(\pi_j ) = \frac{U(\pi_{j}) }{\sum_j U(\pi_{j}) }\; ,
\ee
since all agents have the same objective utilities: $U(\pi_{\alpha j}) = U(\pi_j)$.

The attraction factor, generally, is not the same for different decision
makers (it is not invariant with respect to a change of decision makers)
but, owing to Eqs. (16) and (17), it preserves the {\it alternation conditions}
\be
\label{27}
\sum_{j=1}^L q(\pi_j) = 0 \; , \qquad  -1 \leq q(\pi_j) \leq 1 \;   .
\ee

In this way, each prospect is evaluated by the society with respect to
two characteristics, its utility and its attractiveness. A prospect $\pi_i$ is
more useful than $\pi_j$, if $f(\pi_i) > f(\pi_j)$. And a prospect
$\pi_i$ is more attractive than $\pi_j$, if $q(\pi_i) > q(\pi_j)$.
Therefore, a prospect can be more useful, but not preferred, being less
attractive. As follows from expression (22), a prospect $\pi_1$ is
preferred to a prospect $\pi_2$, in the sense of definition (18), when
\be
\label{28}
  p(\pi_{1}) >  p(\pi_{2}) \qquad ( \pi_{1} > \pi_{2})\;   ,
\ee
if and only if the inequality
\be
\label{29}
f(\pi_{1}) -  f(\pi_{2})   >  q(\pi_{2}) -  q(\pi_{1})
\ee
holds true.

Actually, the comparison of theory with experiment is meaningful only
for a sufficiently large pool of decision makers, when the general
typical features can be defined. In such a large society, when the
number of agents choosing a prospect $\pi_j$ is $N_j$, then the
experimentally observed fraction
\be
\label{30}
p_{exp}(\pi_j) \equiv \frac{N_j}{N}
\ee
provides the aggregate frequentist definition of probability that
should be compared with the theoretical value (22).

For comparing the empirical frequentist probability $p_{exp}(\pi)$
with the theoretical probability $p(\pi)$, we need to know how the
latter can be calculated. The utility factor is explicitly defined
in Eq. (26). The attraction factor, being a subjective quantity,
essentially depends on the subjective state of a decision maker.
Moreover, the same prospect, at different times, can be appreciated
by a decision maker differently. Therefore, it seems that it is so
much random that there is no way of finding its quantitative
definition. However, as is explained above, the attraction factor
possesses some general well defined and fixed properties. For instance,
we know that it varies in the interval $[-1,1]$ and that it obeys the
alternation condition (27). Being a random quantity does not preclude
that it can enjoy some general typical properties. That is, an aggregate
value of the attraction factor can be well defined. Under the aggregate
value, we mean an average value, averaged either over many realizations
of the same problem for a single decision maker or over the results for
many decision makers deciding on a given problem. Such a typical value
of the attraction factor can be found by accomplishing a series of
experimental observations. Another way of theoretically defining the
typical attraction factor is explained in the following section.

\subsection{Typical values of attraction factors}

The attraction factors are subjective quantities that can be different
for different decision makers. And for the same decision maker,
attraction factors are different for different prospects, and even can be
different for the same prospect at different times. This is equivalent to
accepting that the attraction factor is a random quantity that can be
characterized by a distribution $\varphi(q(\pi_{\alpha j}))$. Since the
attraction factor lies in the interval $[-1,1]$, its distribution is
normalized as
\be
\label{A1}
\int_{-1}^1 \vp(q) \; dq =  1 \;   .
\ee
And, in view of the alternation condition (27), the mean value of the
attraction factor is zero,
\be
\label{A2}
 \int_{-1}^1 \vp(q) q\; dq =  0 \;  .
\ee

The exact attraction-factor distribution is unknown in general.
In particular cases, it could be extracted from empirical observations.
Moreover, even in the absence of any a priori empirical information, the
typical values of the attraction factor, being a random quantity varying in
the interval $[-1,1]$, can be estimated \cite{36} in the following way.

Let us define the values
\be
\label{A3}
q_+ \equiv  \int_0^1 \vp(q) q\; dq \; , \qquad
q_- \equiv  \int_{-1}^0 \vp(q) q\; dq \; ,
\ee
which, according to the alternation condition (\ref{A2}), are related as
\be
\label{A4}
 q_+ + q_-  = 0 \;  .
\ee
The absence of any a priori information implies that the distribution
$\varphi (q)$ is uniform. This is evident from the generally accepted notion
of no-a-priori information that implies the equiprobability of the variable
in its whole domain. Also, as is well known, the equiprobable distribution
provides the maximum of the Shannon entropy, which, in turn, characterizes the
information measure \cite{Shannon}.

In the case of the equiprobable distribution, the normalization condition
(\ref{A1}) yields $\varphi (q) = 1/2$. As a result, the values (\ref{A3})
become
\be
\label{A5}
q_+ = \frac{1}{4} \; , \qquad q_- = -\; \frac{1}{4} \;  .
\ee
We have called the existence of such typical values of the attraction factor,
corresponding to the non-informative priors, as the {\it quarter law} \cite{36}.
These values (\ref{A5}) can be used for estimating the influence of the attraction
factors on the decision making of typical agents.

It has been proved in the earlier publications \cite{31,32,35,36} that the quarter
law is in perfect agreement with a variety of empirical observations. Below we also
show that the use of this typical value for the attraction factor is in good
agreement with many other empirical data.

It is worth emphasizing that the value $1/4$ for the typical attraction factor
is valid not only in the case of an equiprobable distribution, but also for a
wide class of distributions. Let us take, for example, the symmetric
beta-distribution
$$
 \vp(q) = \frac{\Gm(2\al)}{2\Gm^2(\al)} \; | q |^{\al-1}
( 1 - | q | )^{\al-1} \;  ,
$$
with the domain $[-1,1]$, often employed in many applications \cite{Abramowitz},
where $\alpha$ is an arbitrary positive parameter. Then the typical values $q_-$
and $q_+$ are exactly $-1/4$ and $+1/4$, respectively, for any $\alpha > 0$.
The same quarter law follows from several other distributions normalized on
the interval $[-1,1]$, for instance, from the symmetric quadratic distribution
$$
\vp(q) = 6 \left ( | q | - \; \frac{1}{2} \right )^2
$$
and from the symmetric triangular distribution
\begin{eqnarray}
\nonumber
\vp(q) = \left \{ \begin{array}{ll}
2 | q | , ~ & ~ 0 \leq | q | \leq \frac{1}{2} \\
2 (1 - | q |) , ~ & ~  \frac{1}{2} <  | q | \leq  1
\end{array} \right. \; .
\end{eqnarray}

In this way, the quarter law is not an ad hoc assumption, but it is a
consequence of the theoretical evaluation for several distributions, which is
in agreement with a number of empirical data.

\subsection{Quantitative resolution of classical paradoxes}

The typical values of the attraction factor (\ref{A5}) make it possible to give
{\it quantitative} predictions for decisions of typical decision makers. For
instance, the disjunction effect, studied in different forms in a variety of
experiments \cite{39}, was thoroughly analyzed \cite{32,36}, and we found that
the empirically determined absolute value of the aggregate attraction factor
$|q(\pi_j)|$ coincided with the value $0.25$ predicted by expressions (\ref{A5}),
within the typical statistical error of the order of $20\%$ characterizing
these experiments. The same quantitative agreement, between the QDT prediction
for the absolute value of the attraction factors and empirical values, holds
for experiments testing the conjunction fallacy \cite{32,36}. The planning paradox
has also found a natural explanation within QDT \cite{31}. Moreover, it has been
shown \cite{35} that QDT explains practically all typical paradoxes of classical
decision making, arising when decisions are taken by typical decision makers.

In order to illustrate how QDT resolves classical paradoxes, let us consider
a typical paradox happening in decision making. In game theory, there is a
series of games in which several subjects can choose either to cooperate with
each other or to defect. Such setups have the general name of {\it prisoner
dilemma games}. The cooperation paradox consists in the real behavior of game
participants who often incline to cooperate despite the prescription of utility
theory for defection. Below, we show that this paradox is easily resolved within QDT,
which gives correct {\it quantitative predictions}.

The generic structure of the prisoner dilemma game is as follows. Two
participants can either cooperate with each other or defect from cooperation.
Let the cooperation action of one of them be denoted by $C_1$ and the
defection by $D_1$. Similarly, the cooperation of the second subject is
denoted by $C_2$ and the defection by $D_2$. Depending on their actions, the
participants receive payoffs from the set
\be
\label{B1}
\mathbb{X} = \{ x_1,x_2,x_3,x_4 \} \;   ,
\ee
whose values are arranged according to the inequality
\be
\label{B2}
 x_3 > x_1 > x_4 > x_2 \;  .
\ee
There are four admissible cases: both participants cooperate $(C_1 C_2)$,
one cooperates and the other defects $(C_1 D_2)$, the first defects but the
second cooperates $(D_1 C_2)$, and both defect $(D_1 D_2)$. The payoffs to each
of them, depending on their actions, are given according to the rule
\begin{eqnarray}
\left [
\begin{array}{cc}
C_1C_2 & C_1D_2 \\
D_1C_2 & D_1D_2 \end{array} \right ] \ra
\left [
\begin{array}{cc}
x_1x_1 & x_2x_3 \\
x_3x_2 & x_4x_4 \end{array} \right ] \; .
\end{eqnarray}
As is clear, the enumeration of the participants is arbitrary, so that it
is possible to analyze the actions of any of them.

Each subject has to decide what to do, to cooperate or to defect, when he/she
is not aware about the choice of the opponent. Then, for each of the
participants, there are two prospects, either to cooperate,
\be
\label{B3}
 \pi_1 = C_1 (C_2 + D_2) \;   ,
\ee
or to defect,
\be
\label{B4}
 \pi_2 = D_1 (C_2 + D_2) \;  .
\ee
The sum $(C_2 + D_2)$ embodies the fact that the decision maker
does not know the choice (cooperate or defect) of the second participant.
In the absence of any information on the action chosen by the opponent,
the probability for each of these actions is $1/2$. Assuming for simplicity a
linear utility function of the payoffs, the expected utility of
cooperation for the first subject is
\be
\label{B5}
 U(\pi_1) = \frac{1}{2} \; x_1 + \frac{1}{2} \; x_2 \;  ,
\ee
while the utility of defection is
\be
\label{B6}
U(\pi_2) = \frac{1}{2} \; x_3 + \frac{1}{2} \; x_4 \;    .
\ee
The assumption of linear utility is not crucial, and can be removed
by reinterpreting the payoff set (\ref{B1}) as a utility set.
Because of condition (\ref{B2}), the utility of defection is always larger
than that of cooperation, $U(\pi_2) > U(\pi_1)$. According to utility theory,
this means that all subjects have always to prefer defection.

However, numerous empirical studies demonstrate that an essential fraction of
participants choose to cooperate despite the prescription of utility theory.
This contradiction between reality and the theoretical prescription constitutes
the cooperation paradox \cite{40,41}.

Considering the same game within the framework of QDT, we have the probabilities
of the two prospects,
\be
\label{B7}
p(\pi_1) = f(\pi_1) + q(\pi_1) \; , \qquad
p(\pi_2) = f(\pi_2) + q(\pi_2) \;   .
\ee

Let us recall that humans possess the so-called propensity for cooperation,
which is a well established empirical fact \cite{Tonnies,Axelrod,Triandis}.
This propensity has developed during the history of humanity starting from
the very beginning of human existence as hunters-gatherers. In the process
of their development, humans noticed that cooperation is profitable for
their survival and well-being. The propensity for cooperation has been the
driving force for the creation of human societies, from tribes to states
and country unions \cite{Tonnies,Axelrod,Triandis,Burke}. Without this
feature, no social groups would be formed. The propensity to cooperation
proposes that the attraction factor for cooperative prospect is larger than
that for the defecting prospect, that is, $q(\pi_1) > q(\pi_2)$. In view of
the alternation law (\ref{27}), we have $q(\pi_1) = - q(\pi_2)$, which can
be estimated by the typical value $1/4$, as in expressions (\ref{A5}).
Hence, we can estimate the considered prospects by the equations
\be
\label{B8}
p(\pi_1) = f(\pi_1) + 0.25 \; , \qquad
 p(\pi_2) = f(\pi_2) - 0.25 \;  .
\ee
From here, we see that, even if defection seems to be more useful than
cooperation, so that $f(\pi_2) > f(\pi_1)$, the cooperative prospect can
be preferred by some of the participants.

To illustrate numerically how this paradox is resolved, let us take the data
from the experimental realization of the prisoner dilemma game by Tversky
and Shafir \cite{39}. Subjects played a series of prisoner dilemma games,
without feedback. Three types of setups were used: first, when the subjects
knew that the opponent had defected; second, when they knew that the opponent
had cooperated; and third, when subjects did not know whether their opponent
had cooperated or defected. The rate of cooperation was $3\%$ when subjects
knew that the opponent had defected, and $16\%$ when they knew that the
opponent had cooperated. However, when subjects did not know whether their
opponent had cooperated or defected, the rate of cooperation was $37\%$.

Treating the utility factors as classical probabilities, we have
$$
f(\pi_1) = \frac{1}{2}\; f(C_1|C_2) + \frac{1}{2}\; f(C_1|D_2) \; ,
$$
$$
f(\pi_2) = \frac{1}{2}\; f(D_1|C_2) + \frac{1}{2}\; f(D_1|D_2) \;   .
$$
According to the Tversky-Shafir data,
$$
f(C_1|C_2) = 0.16 \; , \qquad f(C_1|D_2) = 0.03 \;  .
$$
Therefore,
\be
\label{B9}
 f(\pi_1) = 0.10 \; , \qquad f(\pi_2) = 0.90 \;  .
\ee
Then, for the prospect probabilities (\ref{22}), we get
\be
\label{B10}
p(\pi_1) = 0.35 \; , \qquad p(\pi_2) = 0.65 \;   .
\ee

In this way, the fraction of subjects choosing cooperation is predicted to be
$35 \%$. This is in remarkable agreement with the empirical data of $37 \%$
by \cite{39}. Actually, within the statistical accuracy of the experiment,
the predicted and empirical numbers are indistinguishable.

\section{Influencing choice by reversing attraction factors}

In the prospect probability (22), the first term (23) is an objectively
defined quantity characterizing, depending on the setup, either
a classical probability or the prospect utility factor. It would, of course,
be possible to change the society choice by varying the utility of prospects.
This, however, would be just an objective shift of preferences caused by the
varying prospect utilities.

More important is that it is possible to essentially change the decision
makers choice merely by influencing the attractiveness of the considered
prospects, without essentially varying their utilities. This means that the
attraction factors are to be influenced.

\subsection{Prospect probabilities for binary lattices}

The most often and illustrative case is the choice between two
prospects forming a binary lattice
\be
\label{31}
\cL = \{ \pi_1 , \pi_2 \} \;   .
\ee
Suppose that the prospect $\pi_1$ is more attractive than $\pi_2$,
which means that $q(\pi_1) >  q(\pi_2)$. According to the alternation
property (27), we have $q(\pi_1) = - q(\pi_2)$. Then, taking into account
the quarter law (\ref{A5}), we can estimate the attraction factor
$q(\pi_1)$ as $1/4$, while the attraction factor $q(\pi_2)$ as $-1/4$.
Keeping in mind that a probability, by its meaning, lies in the interval
$[0,1]$, the prospect probabilities can be evaluated by the formulas
$$
p(\pi_1) =
{\rm Ret}_{[0,1]}\left \{ f(\pi_1) + \frac{1}{4} \right \} \; ,
$$
\be
\label{33}
p(\pi_2) =
{\rm Ret}_{[0,1]}\left \{ f(\pi_2) - \frac{1}{4} \right \} \;  ,
\ee
where the retract function
\begin{eqnarray}
\nonumber
{\rm Ret}_{[0,1]}\{ z \} \equiv \left \{
\begin{array}{ll}
0 , & ~ z < 0 \\
z , & ~ 0 \leq z \leq 1 \\
1 , & ~ z > 1 \end{array} \right.
\end{eqnarray}
is employed.

\subsection{Attraction factors and risk aversion}

Formulas (\ref{33}) can be used for evaluating the prospect probabilities
in the case of the binary lattice (\ref{31}). The classification of prospects
onto more or less attractive is based on subjective feelings of decision makers.
Among these, a very important role is played by the notion of aversion to
uncertainty and risk, or ambiguity aversion \cite{42,43,44,45,46,47,48,49}.
It is possible to define as more attractive the prospect that provides more
certain gain, hence, more uncertain loss \cite{36}.

It is worth recalling that the attraction factors are not fixed by the
values $\pm 1/4$. These values have been obtained as non-informative priors
allowing us to estimate the probability of selecting between the prospects.
In particular cases, they can be different, since by their definition, they
characterize subjective features of decision makers. Nevertheless, these
non-informative priors provide a simple way for the probability estimation
and lead to a very good agreement with empirical observations, as has been
shown in our previous publications {\it quantitatively} resolving the classical
paradoxes in decision making, such as disjunction effect and conjunction
fallacy \cite{32,34,36}. And in Sec. III C above, we have shown how the
prisoner-dilemma paradox is {\it quantitatively} resolved within QDT.
Below, we illustrate the applicability of the approach to several examples
considered earlier by Kahneman and Tversky \cite{42}. We stress that our
theory has not been specially designed for explaining these examples, but
the latter provide just one more illustration of the QDT approach that is
general and can be applied to arbitrary cases as has been shown in our previous
publications. We consider below different prospects with gains. The case of
losses is also treatable by QDT. However, this case requires a separate
consideration that is out of the scope of the present paper.

\subsection{Rule for defining attraction factor signs}

The attraction factor sign is principally important, since it essentially
influences the value of the prospect probability. The choice of this sign
depends on the balance between the possible gain and related risk. Below,
we describe how this choice can be done for the most often considered case
of the binary prospect lattice.

Mathematically, the attraction factor, due to mode interference, arises only
for the entangled composite prospects \cite{YS}, which implies decisions under
uncertainty. To the first glance, it seems that deciding between two simple
lotteries does not explicitly involve uncertainty. However, it is necessary
to stress that practically all decisions always deal with uncertainty, though
it may be not explicitly formulated. Suppose, e.g., one has to decide between
two lotteries $\pi_1$ and $\pi_2$. In the process of decision making, the
uncertainty comes from two origins. One is related to the doubt about the
objectivity of the setup suggesting the choice. The other, probably more
important, is the uncertainty caused by the subjective hesitations of the
decision maker with respect to his/her correct understanding of the problem
and his/her knowledge of what would be the best criterion for making a
particular choice. Therefore, even when one formally deals with two simple
lotteries $\pi_1$ and $\pi_2$, one actually confronts the composite prospects
$\pi_1 \bigotimes B$ and $\pi_2 \bigotimes B$, with $B = \{B_1, B_2\}$ being
a set of two events. One of them, $B_1$, represents the confidence of the
decision maker in the empirical setup as well as in the correctness of
his/her decision. The other, $B_2$, corresponds to the disbelief of the
decision maker in the suggested setup and/or in his/her understanding of the
appropriate criteria for the choice. In what follows, we shall keep in mind
the composite prospects $\pi_i \bigotimes B$, with $i = 1,2$, while, for
brevity, we shall write just $\pi_i$.

Let us consider two prospects
$$
 \pi_1 = \{ x_i , p_1(x_i) : \; i = 1,2,\ldots \} \; ,
$$
\be
\label{S1}
\pi_2 = \{ y_j , p_2(y_j) : \; j = 1,2,\ldots \} \;  .
\ee
The related maximal and minimal gains are denoted as
$$
x_{max} \equiv \sup_i \{ x_i \} \; , \qquad
x_{min} \equiv \inf_i \{ x_i \} \; ,
$$
\be
\label{S2}
y_{max} \equiv \sup_j \{ y_j \} \; , \qquad
y_{min} \equiv \inf_j \{ y_j \} \;  .
\ee
The signs of the attraction factors for the binary prospect lattice,
in view of the alternation condition (27), are connected with each other,
$$
 {\rm sgn}\; q(\pi_1) = - {\rm sgn}\; q(\pi_2) \;   ,
$$
because of which in what follows it is sufficient to analyze only one of
them, say the sign of $q(\pi_1)$.

The first prospect gain factor is the ratio
$$
g(\pi_1) \equiv \frac{x_{max}}{y_{max}}
$$
showing how much the maximal gain of the first prospect is larger than
that of the second one. On the other hand, the larger the probability of
getting the minimal gain in the second prospect, the larger is the ratio
$$
r(\pi_2) \equiv \frac{p_2(y_{min})}{p_1(x_{min})} \;  ,
$$
playing the role of the risk factor when choosing the second prospect. The
combined influence of possible gain and risk is described by the product
$g(\pi_1) r(\pi_2)$. The attractiveness of a prospect is characterized by
how much the gain prevails over risk, that is, how the latter product
$g(\pi_1) r(\pi_2)$ differs from one, hence by the sign of the value
\be
\label{S3}
 \al(\pi_1) \equiv g(\pi_1) r(\pi_2) -1 =
\frac{x_{max}p_2(y_{min})}{y_{max}p_1(x_{min})} \; - \; 1 \;  .
\ee
Then the sign of the first prospect attraction factor is defined by the rule
\begin{eqnarray}
\label{S4}
{\rm sgn}\; q(\pi_1) = \left \{
\begin{array}{ll}
+1 , ~ & ~ \al(\pi_1) > 0 \\
-1 , ~ & ~ \al(\pi_1) \leq 0
\end{array}
\right. \;  .
\end{eqnarray}

In the following section, we illustrate the practical application of this
rule and show that it yields the results in good agreement with empirical
observations. Let us stress that the formulated rule is designed for the
case where the utilities of two prospects are close to each other and may
be not applicable when these utilities are strongly different.

\section{Illustration of preference reversal by examples}

Here, we consider several examples of experiments described by Kahneman
and Tversky  \cite{42}. In these experiments, the total number of decision
makers was about equal to or smaller than one hundred, and the corresponding
statistical errors were close to $\pm 0.1$. Decision makers had to
choose between two prospects having the properties as those discussed
above. Payoff were counted in monetary units, say in thousands of
schekels, francs, or dollars. The kind of monetary units has no
influence on the relative quantities, such as utility factors and
prospect probabilities. Calculating the utility factors, we use for
simplicity a linear utility function.

\vskip 2mm

{\bf Example 1}. One chooses between the prospects
$$
\pi_1=\{ 2.5, 0.33\; | \; 2.4, 0.66 \;|\; 0,0.01 \} \; , \qquad
\pi_2=\{ 2.4, 1 \} \;  .
$$
Here, the first number of each pair corresponds to the payoff
while the second number is the associated probability. Thus,
the second prospect $\pi_2$ corresponds to the sure gain (probability $1$)
of $2.4$ monetary units.

Using definition (\ref{26}) of the utility factors
$$
f(\pi_1) = {U(\pi_1) \over U(\pi_1) + U(\pi_2)}~, \qquad
f(\pi_2) = {U(\pi_2) \over U(\pi_1) + U(\pi_2)}~,
$$
with the utilities
$$
U(\pi_1) = 2.5 \times 0.33 + 2.4 \times 0.66 = 2.409 ~, \qquad U(\pi_2)= 2.4~,
$$
gives the utility factors
$$
 f(\pi_1) = 0.501 , \qquad f(\pi_2) = 0.499 \;  .
$$

Following the rule described above, we get
$$
g(\pi_1)=1.042 \; , \qquad r(\pi_2) = 0 \; , \qquad
\al(\pi_1) = -1 \;   ,
$$
which tells us that the first prospect is less attractive. Then, again
employing the non-informative priors, $q(\pi_1)$ can be estimated as $- 1/4$,
while $q(\pi_2)$, as $1/4$. Thus, we get the prospect probabilities
$$
 p(\pi_1) = 0.251 , \qquad p(\pi_2) = 0.749 \;  .
$$
In experiments, it was found that
$$
p_{exp}(\pi_1) = 0.18 , \qquad p_{exp}(\pi_2) = 0.82 \; ,
$$
which, within the experimental accuracy, coincides with the theoretical
prediction.

\vskip 2mm

{\bf Example 2}. One considers the prospects
$$
\pi_1=\{ 2.5, 0.33\; | \; 0, 0.67  \} \; , \qquad
\pi_2=\{ 2.4, 0.34 \; | \; 0, 0.66 \} \;    .
$$
The utility factors are practically the same as in the previous example:
$$
f(\pi_1) = 0.503 , \qquad f(\pi_2) = 0.497 \;   .
$$
Now, the uncertainties of the two prospects are close to each other and
the gain in the first prospect is a bit larger, which gives
$$
 g(\pi_1)=1.042 \; , \qquad r(\pi_2) = 0.985 \; , \qquad
\al(\pi_1) = 0.027 \;   ,
$$
which suggests that the first prospect is more attractive. This yields
the prospect probabilities
$$
  p(\pi_1) = 0.753 , \qquad p(\pi_2) = 0.247 \;  .
$$
Again, this is in agreement with the experimental values
$$
p_{exp}(\pi_1) = 0.83 , \qquad p_{exp}(\pi_2) = 0.17 \; ,
$$
being in the corridor of statistical errors.

Comparing the examples 1 and 2, we see that a change in the distribution
of payoff weights, under the same payoffs, has lead to the reversal of
the attraction factors and, as a result, to the preference reversal.

\vskip 2mm

{\bf Example 3}. The prospects are
$$
 \pi_1=\{ 4, 0.8\; | \; 0, 0.2  \} \; , \qquad
\pi_2=\{ 3, 1 \} \;  .
$$
The utility factors (26) become
$$
 f(\pi_1) = 0.516 , \qquad f(\pi_2) = 0.484 \;  .
$$
In so far as
$$
 g(\pi_1)=1.333 \; , \qquad r(\pi_2) = 0 \; , \qquad
\al(\pi_1) = -1 \;   ,
$$
the second prospect is more attractive. Then, we have the prospect
probabilities
$$
 p(\pi_1) = 0.266 , \qquad p(\pi_2) = 0.734 \;   ,
$$
which agree well with the empirical results
$$
p_{exp}(\pi_1) = 0.2 , \qquad p_{exp}(\pi_2) = 0.8 \;   .
$$

\vskip 2mm

{\bf Example 4}. The prospects
$$
\pi_1=\{ 4, 0.2\; | \; 0, 0.8  \} \; , \qquad
\pi_2=\{ 3, 0.25 \; | \; 0, 0.75 \}
$$
have the same payoffs and the same utility factors
$$
f(\pi_1) = 0.516 , \qquad f(\pi_2) = 0.484 \; ,
$$
as in the previous case. Now we have
$$
g(\pi_1)=1.333 \; , \qquad r(\pi_2) = 0.937 \; , \qquad
\al(\pi_1) = 0.249 \;    .
$$
Hence the first prospect is more attractive. This gives the prospect
probabilities
$$
 p(\pi_1) = 0.766 , \qquad p(\pi_2) = 0.234 \;   ,
$$
with the reverse preference, as compared to Example 3. The experimental
results
$$
p_{exp}(\pi_1) = 0.65 , \qquad p_{exp}(\pi_2) = 0.35
$$
are in agreement with the theoretical prediction.

\vskip 2mm

{\bf Example 5}. For the prospects
$$
 \pi_1=\{ 6, 0.45\; | \; 0, 0.55  \} \; , \qquad
\pi_2=\{ 3, 0.9 \; | \; 0, 0.1 \} \;   ,
$$
the utility factors are equal,
$$
 f(\pi_1) = 0.5 , \qquad f(\pi_2) = 0.5 \;  .
$$
The second prospect is more attractive, since
$$
g(\pi_1)=2 \; , \qquad r(\pi_2) = 0.182 \; , \qquad
\al(\pi_1) = -0.636 \;    .
$$
Then
$$
  p(\pi_1) = 0.25 , \qquad p(\pi_2) = 0.75 \;  .
$$
The experimental results
$$
p_{exp}(\pi_1) = 0.14 , \qquad p_{exp}(\pi_2) = 0.86 \; ,
$$
within the statistical errors of $\pm 0.1$, agree with the theoretical
prediction.

\vskip 2mm

{\bf Example 6}. The prospects
$$
\pi_1=\{ 6, 0.001\; | \; 0, 0.999  \} \; , \qquad
\pi_2=\{ 3, 0.002 \; | \; 0, 0.998 \}
$$
lead to the same utility factors
$$
f(\pi_1) = 0.5 , \qquad f(\pi_2) = 0.5 \;   ,
$$
as in the previous example. The uncertainties of the two prospects are close
to each other. However, the gain in the first prospect is essentially
larger, which gives
$$
g(\pi_1)=2 \; , \qquad r(\pi_2) = 0.999 \; , \qquad
\al(\pi_1) = 0.998 \;   .
$$
This makes the second prospect less attractive. As a result, the prospect
preference reverses, as compared to Example 5, with the prospect
probabilities
$$
 p(\pi_1) = 0.75 , \qquad p(\pi_2) = 0.25 \;   .
$$
The experimental data
$$
p_{exp}(\pi_1) = 0.73 , \qquad p_{exp}(\pi_2) = 0.27
$$
practically coincide with the theoretical prediction, again demonstrating
the preference reversal.

\vskip 2mm

{\bf Example 7}. Consider the prospects
$$
\pi_1=\{ 6, 0.25\; | \; 0, 0.75  \} \; , \qquad
\pi_2=\{ 4, 0.25 \; | \; 2, 0.25| \; 0, 0.5 \}
$$
The utility factors are
$$
f(\pi_1) = 0.5 , \qquad f(\pi_2) = 0.5 \;   ,
$$
Now we have
$$
 g(\pi_1)=1.5 \; , \qquad r(\pi_2) = 0.667 \; , \qquad
\al(\pi_1) = 0 \;  .
$$
Hence, the first prospect is less attractive, leading to the probabilities
$$
 p(\pi_1) = 0.25 , \qquad p(\pi_2) = 0.75 \;   .
$$
The empirical data
$$
p_{exp}(\pi_1) = 0.18 , \qquad p_{exp}(\pi_2) = 0.82 \; ,
$$
within the experimental accuracy, are in agreement with the theoretical
prediction.

\vskip 2mm

The results are summarized in the Table where, in the last column, the
error
$$
 \Dlt(\pi_1) = | p(\pi_1) - p_{exp}(\pi_1) |
$$
is shown. For all cases, this error is about $0.1$, which is the same
as the standard error $0.1$ for the corresponding experiments.

\begin{table}[!t]
\renewcommand{\arraystretch}{1.5}
\caption{Utility factor $f(\pi_1)$, prospect probability $p(\pi_1)$, empirical
frequency $p_{exp}(\pi_1)$, and the deviation $\Dlt(\pi_1)$ for the
considered examples of decision-making manipulation.}
\centering
\begin{tabular}{|l|l|l|l|l|}  \hline
    & $f(\pi_1)$  & $p(\pi_1)$ & $p_{exp}(\pi_1)$ & $\Dlt(\pi_1)$ \\ \hline
1   & 0.501  & 0.251      & 0.18   & 0.07 \\ \hline
2   & 0.503  & 0.753      & 0.83   & 0.08 \\ \hline
3   & 0.516  & 0.266      & 0.20   & 0.07  \\ \hline
4   & 0.516  & 0.766      & 0.65   & 0.12  \\ \hline
5   & 0.5    & 0.25       & 0.14   & 0.11  \\ \hline
6   & 0.5    & 0.75       & 0.73   & 0.02  \\ \hline
7   & 0.5    & 0.25       & 0.18   & 0.07  \\ \hline
\end{tabular}

\end{table}

In the above examples, we have considered prospects that are characterized
by different gains. The treatment of prospects, involving losses, is a separate
problem. Strictly speaking, in real life, in order to lose something, it is
in general the case that one possesses a wealth no less than the loss. However,
there are also examples of negative wealth, associated with debts that are
larger than present equity. For a firm, this leads in general to bankruptcy.
Rationally, agents should also default on their debts, if they can, a situation
that often but not always occurs, as for instance exemplified by the
many cases of negative equity of homeowners in the USA \cite{50} and Great
Britain \cite{51} following the real estate price collapse and financial
crisis. Thus, in general, we should expect that the prospect probabilities depend
on the initial richness of decision makers. But in the laboratory experiments,
one usually considers artificial situations, with imaginary or unrealistic small
losses, when the starting assets are not important. The real and imaginary losses
are rather different things and are to be treated differently. These delicate
problems are out of the scope of the present paper and will be treated in a
separate publication.

Our aim has been to demonstrate the fact that, under practically the same utility,
by appropriately arranging the payoff weights, it is possible to realize the
reversal of the attraction factors and, as a result, the reversal of decision
preferences.

\section{Influence by varying available information}

The standard setup of studying decision making in the laboratory is when
decision makers are assumed to give responses without consulting each
other and without looking for additional information. However, in a number
of experimental studies, it has been found that decisions can essentially
change when the agents are allowed to consult with each other, increasing
by this their mutual information \cite{52,53,54,55,56,57,58}, or when they
can get additional information by learning from their own experience \cite{59}.

When the objective parts of the prospect probabilities are assumed to remain
invariant, the influence on decision making of the obtained information can be
realized by varying the attraction factors. Therefore, we have to understand
how the latter vary with respect to the change of information available to
decision makers.

Let us denote by $\mu$ the measure of information available to a decision maker.
This measure can be defined according to one of the known ways of measuring
information \cite{60}. Decision making depends on the amount of information and
varies when it changes \cite{61}.

Generalizing the consideration of Sec. II, we take into account that the
society state, represented by the statistical operator $\hat\rho(\mu)$,
depends on the available information $\mu$. This means that the society state
$\hat\rho$ is influenced by the received additional information $\mu$, which
transforms $\hat\rho$ into $\hat\rho(\mu)$. The transformation law can be
represented by a unitary evolution operator, as is described below. In simple
language, this implies that the society state depends on the available
information and varies when the level of information changes \cite{Dixit}.
In mathematical terms, the amount of the received information can be
quantified by the Kullback-Leibler \cite{Kullback} information
$$
 I_{KL}(\mu) = {\rm Tr} \hat \rho(\mu)
\ln \; \frac{\hat\rho(\mu)}{\hat\rho} \;  .
$$

The prospect probabilities of an $\alpha$ - agent take the form
\be
\label{41}
 p(\pi_{\al j},\mu) = {\rm Tr}_\cH \hat\rho(\mu) \hat P(\pi_{\al j})\;,
\ee
where all notations are the same as in Sec. II.

The variation of the society state with information can be described by
the information evolution operator $\hat{U}(\mu)$, so that
\be
\label{42}
 \hat\rho(\mu) =\hat U(\mu) \hat\rho \hat U^+(\mu) \;  ,
\ee
where
\be
\label{43}
  \hat\rho(0) =\hat\rho \; .
\ee
As before, the society state is normalized, such that
\be
\label{44}
 {\rm Tr}_\cH \hat\rho(\mu) = 1\; .
\ee

The initial condition (\ref{43}) yields
\be
\label{45}
\hat U(0) = \hat 1_\cH \;  ,
\ee
with $\hat{1}_\cH$ being the unity operator on space (2).
And the normalization condition (\ref{44}) requires that the evolution
operator be a unitary operator:
\be
\label{46}
 \hat U^+(\mu) \hat U(\mu) = \hat 1_\cH \;  .
\ee
These properties make it possible to represent the evolution
operator as
\be
\label{47}
  \hat U(\mu) = e^{-i\hat H\mu} \; ,
\ee
where $\hat{H}$, acting on space (2), is called the evolution
generator.

The general form of the evolution generator can be written as
the sum of the terms acting on each of the decision makers in the
society and the term characterizing the interactions between these 
decision makers:
\be
\label{48}
 \hat H = \bigoplus_{\al = 1}^N \; \hat H_\al + \hat H_{int} \; ,
\ee
where $\hat{H}_\alpha$ acts on space (1) and $\hat{H}_{int}$,
on space (2).

The agents of the  society are considered as separate individuals
who, though interacting with each other, do not loose their personal
identities and are able to take individual decisions. In mathematical
language, this means that agents are quasi-isolated \cite{62}. The
mathematical formulation of the quasi-isolated state reads as the
commutation condition
\be
\label{49}
 \left [ \hat H_\al \bigotimes  \hat 1_\cH \; , \; \hat H_{int}
\right ] = 0 \;  .
\ee

Similarly to Sec. III, assuming that all agents are confronted with
the same prospect lattice, we introduce the notion of a typical agent,
whose decisions are described by the average prospect probabilities
\be
\label{50}
 p(\pi_j,\mu) = \frac{1}{N} \sum_{\al=1}^N p(\pi_{\al j},\mu) \;  .
\ee
The property of quasi-isolation (\ref{49}) makes it possible to show
that the prospect probabilities (\ref{50})
acquire the form
\be
\label{51}
 p(\pi_{j},\mu) = f(\pi_j) + q(\pi_j,\mu) \;  ,
\ee
similar to Eq. (22). Here, the first term is the utility factor that
is the same as in Eqs. (23) and (26). The second term is the attraction
factor that can be represented as
\be
\label{52}
 q(\pi_j,\mu)  = q(\pi_j) D(\mu) \;   ,
\ee
where
\be
\label{53}
q(\pi_j)  = q(\pi_j,0)
\ee
is the attraction factor at the initial state, when no additional
information has yet been digested, and $D(\mu)$ is a decoherence
factor. This name comes from the fact that, technically, the attraction
factor appears under the interference of composite prospects \cite{36}.
Decoherence implies that the interference effects fade away, so that
the prospect probabilities tend to their classical values defined by
the utility factors. In other words, this means that
\be
\label{54}
   \lim_{\mu\ra\infty} p(\pi_j,\mu) = f(\pi_j) \; .
\ee
Treating the agent interactions in analogy with a scattering process over
random scatterers, with the width $\mu_c$ in the Lorentzian distribution
of scatterer defects \cite{62}, we have
\be
\label{55}
 D(\mu) = \exp\left ( -\; \frac{\mu}{\mu_c} \right ) \;  .
\ee
The meaning of $\mu_c$ is the amount of information required for the
reduction of the attraction factor by a ratio of $e=2.718...$.

The dependence of the attraction factors on the available information
suggests that it is admissible to vary these factors by regulating
the amount of information. Respectively, by varying the attraction
factors, it is possible to influence decisions. For instance, suppose
that, at $\mu = 0$, the prospect $\pi_1$ is preferred to $\pi_2$.
By providing additional information, one can reduce the attraction
factors according to Eq. (\ref{55}). As a result, the preference
can be reversed, with the prospect $\pi_2$ becoming preferable
to $\pi_1$.

In a series of experimental studies, it has been found that decisions
essentially change when the agents are allowed to consult with each
other, increasing in this way their mutual information
\cite{52,53,54,55,56,57,58}, or when they can get additional
information by learning from their own experience \cite{59}. In these
experiments \cite{52,53,54,55,56,57,58,59}, it has been proved that
additional information does decrease the errors in decision making,
which in our notation correspond to the diminishing decoherence factor $D(\mu)$.
However, in those experiments, one considers two-step procedures. The
knowledge of only two points does not allow for defining the whole
function. Therefore more detailed experiments analyzing multi-step
procedures are needed for comparing empirical results with the form
of the theoretical decoherence factor.

Note that it is possible to provide correct information as well as
incorrect one, the latter corresponding to the process of confusing
decision makers and forcing them to accept some desired decision. The
effect, similar to providing negative information, can be achieved if
decision makers are asked to deliberate concentrating of the uncertainty
contained in the considered prospects \cite{63}.

It is worth stressing that, while the attraction factor and hence the
decision makers choice can be influenced by the provided additional
information, decision makers can never become completely rational.
This is because the amount of information cannot be infinite. Therefore,
the attraction factor is never exactly zero.

\section{Conclusion}

We have studied how the choice of decision makers can be influenced
under the presence of risk and uncertainty. Our analysis is based on
the Quantum Decision Theory that has been previously developed by the
authors for individual decision makers. The term ``quantum" does not
imply that decision makers are assumed to be quantum objects, but it
reflects the use of mathematical techniques that are common for quantum
theory, in particular, for the definition of event probabilities. In quantum
theory, these mathematical techniques make it possible to take into
account unknown hidden variables, at the same time, avoiding their explicit
consideration. Similarly, in decision theory, these techniques allow for
taking into account such hidden variables as subconscious feelings, emotions,
and behavioral biases.

We have suggested a generalization of the theory to the case of decision
makers that are members of a society. The social decision makers interact with
each other by exchanging information. The notion of a typical decision maker,
representing the average society behavior, has been introduced and
characterized.

Under the given utility of prospects, the typical behavior of agents
can be influenced. Changing the results of decision making can be realized by
influencing the attraction factor of decision makers. This can be done
in two ways. One method is to arrange the payoff weights so as to
induce the required changes of the attraction factors. The variation
of the payoff weights can invert the attraction-factor values and
reverse the decision preferences. The second method of influencing is
by providing information to decision makers or by allowing consultations
between the agents of the society. The attraction factors can be either
decreased, when decision makers obtain correct information, or increased
if the delivered information is wrong. The variation of the attraction factors,
induced by positive or negative information, can lead to the reversal
of preferences. Since the amount of information is never infinite, the
attraction factors cannot be reduced exactly to zero. This means that decision
makers cannot become absolutely rational and will always exhibit
some behavioral biases.

The possibility of influencing decision makers is, of course, not a
novelty. What is principally new in the present paper is the
{\it mathematical description} of the process allowing for
{\it quantitative predictions}. By treating several concrete decision
problems, we have illustrated that our theory yields theoretical
predictions that, within experimental accuracy, coincide with empirical
results.

In the present paper, we have considered the application of the Quantum
Decision Theory to human decision making. This, however, is only one of
the admissible applications. Having in hands a well developed mathematical
theory, it is possible to apply it to the problem of creating artificial
quantum intelligence \cite{33} and to use it for quantum information
processing \cite{95}. Understanding the logic of functioning of the human
brain can give us hints on the optimal ways of arranging the functioning
of artificial machine devices. In that sense, the Quantum Decision Theory
plays a special role. From one side, it makes it possible to give
unambiguous interpretation of human decision making. And from another side,
this theory can be used for organizing artificial processes imitating the
logic of humans. Some ideas on the feasibility of creating artificial
quantum intelligence have been advanced in Ref. \cite{33} and a model of
quantum information processing has been analyzed in Ref. \cite{95}. The
detailed consideration of such artificial processes is a separate problem
that needs additional investigations.

\section*{Acknowledgment}

We are very grateful to E.P. Yukalova for many useful discussions.
Financial support from the Swiss National Science Foundation is
appreciated.

\ifCLASSOPTIONcaptionsoff
  \newpage
\fi




\begin{thebibliography}{1}

\bibitem{1}
T.D. Wilson and J.W. Schooler,
"Thinking too much: introspection can reduce the quality of preferences
and decisions",
{\it Journal of Personality and Social Psychology},
vol. 60, pp. 181--192, 1991.


\bibitem{2}
T.D. Wilson, D.B. Centerbar, D.A. Kermer, and D.T. Gilbert,
"The pleasures of uncertainty: prolonging positive moods in ways people
do not anticipate",
{\it Journal of Personality and Social Psychology},
vol. 88, pp. 5--21, 2005.


\bibitem{3}
U. Gneezy, J.A. List, and G. Wu,
"The uncertainty effect: when a risky prospect is valued less than its
worst possible outcome",
{\it Quarterly Journal of Economics}, vol. 121, pp. 1283--1309, 2006.


\bibitem{4}
M.I. Norton, J.H. Frost, and D. Ariely,
"Less is more: the lure of ambiguity, or why familiarity breeds contempt",
{\it Journal of Personality and Social Psychology},
vol. 92, pp. 97--105, 2007.


\bibitem{5}
J.W. Payne, A. Samper, J.R. Bettman, and M.F. Luce,
"Boundary conditions on unconscious thought in decision making",
{\it Psychological Science}, vol. 19, pp. 1118--1123, 2008.


\bibitem{6}
J.J. Koehler and G. James,
"Probability matching in choice under uncertainty: intuition versus
deliberation",
{\it Cognition}, vol. 113, pp. 123--127, 2009.

\bibitem{7}
P.E. Polister,
"Cognitive guidelines for simplifying medical information: data framing
and perception",
{\it Journal of Behavioral Decision Making}, vol. 2, pp. 149--165, 1989.

\bibitem{8}
R.E. O'Carrol and  B.P. Papps,
"Decision making in humans: the effect of manipulating the central
noradrenergetic system",
{\it Journal of Neurology Neurosurgery and Psychiatry},
vol. 74, pp. 376--378,  2003.

\bibitem{9}
B.R. Barber,
{\it Strong Democracy}. Berkeley: University of California, 1984.

\bibitem{10}
A. Gutmann and D. Thompson,
{\it Democracy and Disagreement}. Cambridge: Harvard University, 1996.

\bibitem{11}
J.K. Dryzek,
{\it Deliberative Democracy and Beyond}. Oxford: Oxford University, 2000.

\bibitem{12}
S. Chambers,
"Deliberative democratic theory",
{\it Annual Review of Political Science}, vol. 6, pp. 307--326, 2003.

\bibitem{13}
C. Hafer and D. Landa,
"Deliberation as self-discovery",
{\it Journal of Theoretical Politics}, vol. 19, pp. 329--360, 2007.

\bibitem{14}
Y. Schul,  and Y. Ganzach,
"The effects of accessibility of standards and decision framing on product
evaluations",
{\it Journal of Consumer Psychology}, vol. 4, pp. 61--83, 1995.

\bibitem{15}
K.C. Armel, A. Beaumel, and A. Rangel,
"Biasing simple choices by manipulating relative visual attention",
{\it Judgement and Decision Making}, vol. 3, pp. 396--403, 2008.

\bibitem{16}
D. Lerouge,
"Evaluating the benefits of distraction on product evaluation:
the mind-set effect",
{\it Journal of Consumer Research}, vol. 36, pp. 367--379, 2009.

\bibitem{17}
 J.R. Busemeyer and J.T. Townsend,
"Decision field theory: a dynamic cognition approach to decision making",
{\it Psychological Review}, vol. 100, pp. 432--459, 1993.

\bibitem{18}
T.B. Heath, S. Chatterjee, and K.R. France,
"Mental accounting and changes in price: the frame dependence of
reference dependence",
{\it Journal of Consumer Research}, vol. 22, pp. 90--97, 1995.

\bibitem{19}
J.R. Busemeyer and A. Diederich,
"Survey of decision field theory",
{\it Mathematical Social Sciences}, vol. 43, pp. 345--370, 2002.

\bibitem{20}
K.C. Armel and A. Rangel,
"The impact of computation time and experience on decision values",
{\it American Economic Review}, vol. 98, pp. 163--168, 2008.

\bibitem{21}
J.J. Kellaris, R.K. Frank, and T. DiNovo,
"Exploring the boundaries of the framing effect: the moderating roles
of disparate expected values and perceived costs of judgemental errors",
{\it Marketing Letters}, vol. 6, pp. 175--182,1995.

\bibitem{22}
K.L. Keller,
"Cue compatibility and framing in advertising",
{\it Journal of Marketing Research}, vol. 28, pp. 42--57,1991.

\bibitem{23}
L. Lee, O. Amir, and D. Ariely,
"In search of homo economicus: cognitive noise and the role of
emotion in preference consistency",
{\it Journal of Consumer Research}, vol. 36, pp. 173--187, 2009.

\bibitem{24}
K. Goldsmith and O. Amir,
"Can uncertainty improve promotions?",
{\it Journal of Marketing Research}, vol. 47, pp. 1070--1077, 2010.

\bibitem{25}
W.J. Qualls and C.P. Puto,
"Organizational climate and decision framing: an integrated approach
to analyzing industrial buying decisions",
{\it Journal of Marketing Research}, vol. 26, pp. 179--192, 1989.

\bibitem{26}
B.J. Gibbs,
"Predisposing the decision maker versus framing the decision:
a consumer-manipulation approach to dynamic preference",
{\it Marketing Letters}, vol. 8, pp. 71--83, 1997.

\bibitem{27}
J.J. Koehler, B.J. Gibbs, and R.M. Hogarth,
"Shattering the illusion of control: multi-shot versus single-shot
gambles",
{\it Journal of Behavioral Decision Making}, vol. 7, pp. 183--191, 1994.

\bibitem{28}
K.A. Kinsey, H.G. Grasmick, and K.W. Smith,
"Framing justice: taxpayer evaluations of personal tax burdens",
{\it Law and Society Review}, vol. 25, pp. 845--873, 1991.

\bibitem{29}
R.M. Hogarth, B.J. Gibbs, C.R. McKenzie, and M.A. Marquis,
"Learning from feedback: exactingness and incentives",
{\it Journal of Experimental Psychology: Learning Memory and Cognition},
vol. 17, pp. 734--752, 1991.

\bibitem{Neumann_30}
J. von Neumann and O. Morgenstern,
{\it Theory of Games and Economic Behavior}.
Princeton: Princeton University, 1953.

\bibitem{42}
D. Kahneman and A. Tversky,
"Prospect theory: an analysis of decision under risk"
{\it Econometrica}, vol. 47, pp. 263--292, 1979.

\bibitem{43}
A. Tversky and D. Kahneman,
"Advances in prospect theory: cumulative representation of uncertainty",
{\it Journal of Risk and Uncertainty}, vol. 5, pp. 297--323, 1992.

\bibitem{Quiggin_30}
J. Quiggin,
"A theory of anticipated utility",
{\it Journal of Economic Behavior and Organization},
vol. 3, pp. 323--343, 1982.

\bibitem{Gilboa_30}
I. Gilboa,
"Expected utility with purely subjective non-additive probabilities",
{\it Journal of Mathematical Economics}, vol. 16, pp. 65--88, 1987.

\bibitem{Schmeidler_30}
D. Schmeidler,
"Subjective probability and expected utility without additivity",
{\it Econometrica}, vol. 57, pp. 571--587, 1989.

\bibitem{Gilboa_31}
I. Gilboa and D. Schmeidler,
"Maxmin expected utility with non-unique prior",
{\it Journal of Mathematical Economics},
vol. 18, pp. 141--153, 1989.

\bibitem{Cohen_31}
M. Cohen and J.M. Tallon,
"Decision dans le risque et l'incertain: l'apport des modeles non additifs",
{\it Revue d'Economie Politique}, vol. 110, pp. 631--681, 2000.

\bibitem{Montesano_31}
A. Montesano,
"Effects of uncertainty aversion on the call option market",
{\it Theory and Decision}, vol. 65, pp. 97--123, 2008.

\bibitem{40}
C. Camerer,
{\it Behavioral Game Theory}.
Princeton: Princeton University, 2003.

\bibitem{Shefrin_31}
H.M. Shefrin,
{\it A Behavioral Approach to Asset Pricing}.
New York: Academic, 2005.

\bibitem{Guo_31}
P. Guo,
"One-shot decision theory",
{\it IEEE Transactions on Systems, Man, and Cybernetics--Part A:
Systems and Humans}, vol. 41, pp. 917--926, 2011.

\bibitem{Epstein_31}
L.G. Epstein,
"Living with risk",
{\it Review of Economical Studies}, vol. 75, pp. 1121--1141, 2008.

\bibitem{West_31}
B.J. West and P. Grigolini,
"A psychological model of decision making",
{\it Physica A}, vol. 389, pp. 3580--3587, 2010.

\bibitem{Blavatskyy_31}
P. Blavatskyy,
"A simple behavioral characterization of subjective expected utility",
{\it Operations Research}, vol. 61, pp. 932--940, 2013.

\bibitem{Selten_31}
R. Selten,
"Aspiration adaptation theory",
{\it Journal of Mathematical Psychology}, vol. 42, pp. 191--214, 1998.

\bibitem{Napel_31}
S. Napel,
"Aspiration adaptation in the ultimatum minigame",
{\it Games and Economic Behavior}, vol. 43, pp. 86--106, 2003.

\bibitem{Camerer_41}
C.F. Camerer, G. Loewenstein, and R. Rabin (eds.),
{\it Advances in Behavioral Economics}.
Princeton: Princeton University, 2003.

\bibitem{Machina_42}
M.J. Machina,
"Non-expected utility theory", in S.N. Durlauf and L.E. Blume (eds.),
{\it New Palgrave Dictionary of Economics}.
New York: Macmillan, 2008.

\bibitem{Safra_42}
Z. Safra and U. Segal,
"Calibration results for non-expected utility theories",
{\it Econometrica}, vol. 76, pp. 1143--1166, 2008.

\bibitem{Al-Najjar_42}
N.I. Al-Najjar and J. Weinstein,
"The ambiguity aversion literature: a critical assessment",
{\it Economics and Phylosophy}, vol. 25, pp. 249--284, 2009.

\bibitem{Al-Najjar_43}
N.I. Al-Najjar and J. Weinstein,
"Rejoinder: the ambiguity aversion literature: a critical assessment",
{\it Economics and Phylosophy}, vol. 25, pp. 357--369, 2009.

\bibitem{30}
V.I. Yukalov and D. Sornette,
"Quantum decision theory as quantum theory of measurement",
{\it Physics Letters A}, vol. 372, pp. 6867--6871, 2008.

\bibitem{31}
V.I. Yukalov and D. Sornette,
"Physics of risk and uncertainty in quantum decision making",
{\it European Physical Journal B}, vol. 71, pp. 533--548, 2009.

\bibitem{32}
V.I. Yukalov and D. Sornette,
"Processing information in quantum decision theory.",
{\it Entropy}, vol. 11, pp. 1073--1120, 2009.

\bibitem{33}
V.I. Yukalov and D. Sornette,
"Scheme of thinking quantum systems",
{\it Laser Physics Letters}, vol. 6, pp. 833--839, 2009.

\bibitem{34}
V.I. Yukalov and D. Sornette,
"Entanglement production in quantum decision making",
{\it Physics of Atomic Nuclei}, vol. 73, pp. 559--562, 2010.

\bibitem{35}
V.I. Yukalov and D. Sornette,
"Mathematical structure of quantum decision theory",
{\it Advances in Complex Systems}, vol. 13, pp. 659--698, 2010.

\bibitem{36}
V.I. Yukalov and D. Sornette,
"Decision theory with prospect interference and entanglement",
{\it Theory and Decision}, vol. 70, pp. 283--328, 2011.

\bibitem{Frederick_36}
S. Frederick, G. Loewenstein, and T. O'Donoghue,
"Time discounting and time preference: a critical review",
{\it Journal of Economic Literature}, vol. 40, pp. 351--401, 2002.

\bibitem{Rambaud_36}
S.C. Rambaud and M.J.M. Torrecillas,
"Some considerations on the social discount rate",
{\it Environmental Science and Policy}, vol. 8, pp. 343--355, 2005.

\bibitem{Dirac}
P.A.M. Dirac,
{\it The Principles of Quantum Mechanics},
Oxford: Clarendon, 1958.

\bibitem{38}
J. von Neumann,
{\it Mathematical Foundations of Quantum Mechanics}.
Princeton: Princeton University, 1955.

\bibitem{YS}
V.I. Yukalov and D. Sornette,
"Quantum probabilities of composite events in quantum measurements with
multimode states",
{\it Laser Physics}, vol. 23, p. 105502, 2013.

\bibitem{Loewenstein}
G. Loewenstein,
"Out of control: visceral influences on behavior",
{\it Organizational Behavior and Human Decision Processes},
vol. 65, pp. 272--292, 1996.

\bibitem{Shannon}
C.E. Shannon,
"A mathematical theory of communication",
{\it Bell System Technical Journal}, vol. 27, pp. 379--423, 1948.

\bibitem{Abramowitz}
M. Abramowitz and I.A. Stegun,
{\it Handbook of Mathematical Functions}.
New York: Dover, 1970.

\bibitem{39}
A. Tversky and E. Shafir,
"The disjunction effect in choice under uncertainty",
{\it Psychological Sciences}, vol. 3, pp. 305--309, 1992.

\bibitem{41}
A. Tversky,
{\it Preference, Belief, and Similarity: Selected Writings}.
Cambridge: Massachusetts Institute of Technology, 2004.

\bibitem{Tonnies}
F. T\"{o}nnies,
{\it Community and Association}. New York: Harper Torchbook, 1957.

\bibitem{Axelrod}
R. Axelrod,
{\it The Evolution of Cooperation}. New York: Basic Books, 1984.

\bibitem{Triandis}
H.C. Triandis,
{\it Individualism and Collectivism}. Boulder: Westview Press, 1995.

\bibitem{Burke}
E. Burke,
{\it Reflections on the Revolution in France}. New York: Anchor Press, 1973.

\bibitem{44}
G. Gollier,
{\it Economics of Risk and Time}. Cambridge: Massachusetts Institute of
Technology, 2001.

\bibitem{45}
D. Sornette,
{\it Why Stock Markets Crash}. Princeton: Princeton University. 2003.

\bibitem{46}
Y. Malevergne and D. Sornette,
{\it Extreme Financial Risks}. Heidelberg: Springer, 2006.

\bibitem{47}
G. van de Kuilen,
"Subjective probability weighting and the discovered preference hypothesis",
{\it Theory and Decision}, vol. 67, pp. 1--22, 2009.

\bibitem{48}
M. Abdellaoui, A. Baillon, L. Placido, and P.P. Wakker,
"The rich domain of uncertainty: source functions and their
experimental implementation",
{\it American Economic Review}, vol. 101, pp. 695--723, 2011.

\bibitem{49}
M. Abdellaoui, A. Driouchi, and O. L'Haridon,
"Risk aversion elicitation: reconciling tractability and bias
minimization",
{\it Theory and Decision}, vol. 71, pp. 63--80, 2011.

\bibitem{50}
N. Bhutta, H. Shan, and J.K. Dokko,
"The depth of negative equity and mortgage default decisions",
{\it FEDS Working Paper}, no. 2010-35, 2010.
http://ssrn.com/abstract=1895493,

\bibitem{51}
T. Hellebrandt, S. Kawar, and M. Waldron,
"The economics and estimation of negative equity.
Bank of England Quarterly Bulletin 2009 Q2", 2009.
http://ssrn.com/abstract=1420033.

\bibitem{52}
G. Charness and M. Rabin,
"Understanding social preferences with simple tests",
{\it Quarterly Journal of Economics}, vol. 117, pp. 817--869,  2002.

\bibitem{53}
A. Blinder and J. Morgan,
"Are two heads better than one? An experimental analysis of group versus
individual decision-making",
{\it Journal of Money Credit Banking}, vol. 37, pp. 789--811, 2005.

\bibitem{54}
D. Cooper and J. Kagel,
"Are two heads better than one? Team versus individual play in signaling
games",
{\it American Economic Review}, vol. 95, pp.  477--509, 2005.

\bibitem{55}
G. Charness, E. Karni, and D. Levin,
"Individual and group decision making under risk: an experimental study
of Bayesian updating and violations of first-order stochastic dominance",
{\it Journal of Risk and Uncertainty}, vol. 35, pp. 129--148, 2007.

\bibitem{56}
G. Charness, L. Rigotti, and A. Rustichini,
"Individual behavior and group membership",
{\it American Economic Review}, vol. 97, pp. 1340--1352, 2007.

\bibitem{57}
Y. Chen and S. Li,
"Group identity and social preferences",
{\it American Economic Review}, vol. 99, pp. 431--457, 2009.

\bibitem{58}
G. Charness, E. Karni, and D. Levin,
"On the conjunction fallacy in probability judgement:
new experimental evidence regarding Linda",
{\it Games and Economic Behavior}, vol. 68, pp. 551--556, 2010.

\bibitem{59}
A. K\"{u}hberger, D. Komunska, and J. Perner,
"The disjunction effect: does it exist for two-step gambles?",
{\it Organizational Behavior and Human Decision Processes},
vol. 85, pp. 250--264, 2001.

\bibitem{60}
C. Arndt,
{\it Information Measures}. Berlin: Springer, 2004.

\bibitem{61}
D. Dong, C. Chen, H. Li, and T.J. Tarn,
"Quantum reinforcement learning",
{\it IEEE Transactions on Systems, Man, and Cybernetics B},
vol. 38, pp. 1207--1220, 2008.

\bibitem{Dixit}
A. Dixit and T. Besley,
"James Mirrlees contribution to the theory of information and incentives",
{\it Scandinavian Journal of Economics}, vol. 99, pp. 207--235, 1997.

\bibitem{Kullback}
S. Kullback, (1959)
{\it Information Theory and Statistics}, New York: John Wiley, 1959.

\bibitem{62}
V.I. Yukalov,
"Equilibration of quasi-isolated quantum systems",
{\it Physics Letters A}, vol. 376, pp. 550--554, 2012.

\bibitem{63}
L. Waroquier, D. Marchiori, O. Klein, and A. Cleeremans,
"Is it better to think unconsciously or to trust your first impression?",
{\it Social Psychological and Personality Science} vol. 1, pp. 111--118, 2010.

\bibitem{95}
V.I. Yukalov, E.P. Yukalova, and D. Sornette,
"Mode interference in quantum joint probabilities for multimode Bose-condensed
systems", {\it Laser Physics Letters} vol. 10, p. 115502, 2013.

\end{thebibliography}
%

%

\begin{IEEEbiography}[{\includegraphics[width=1in,height=1.25in,clip,keepaspectratio]{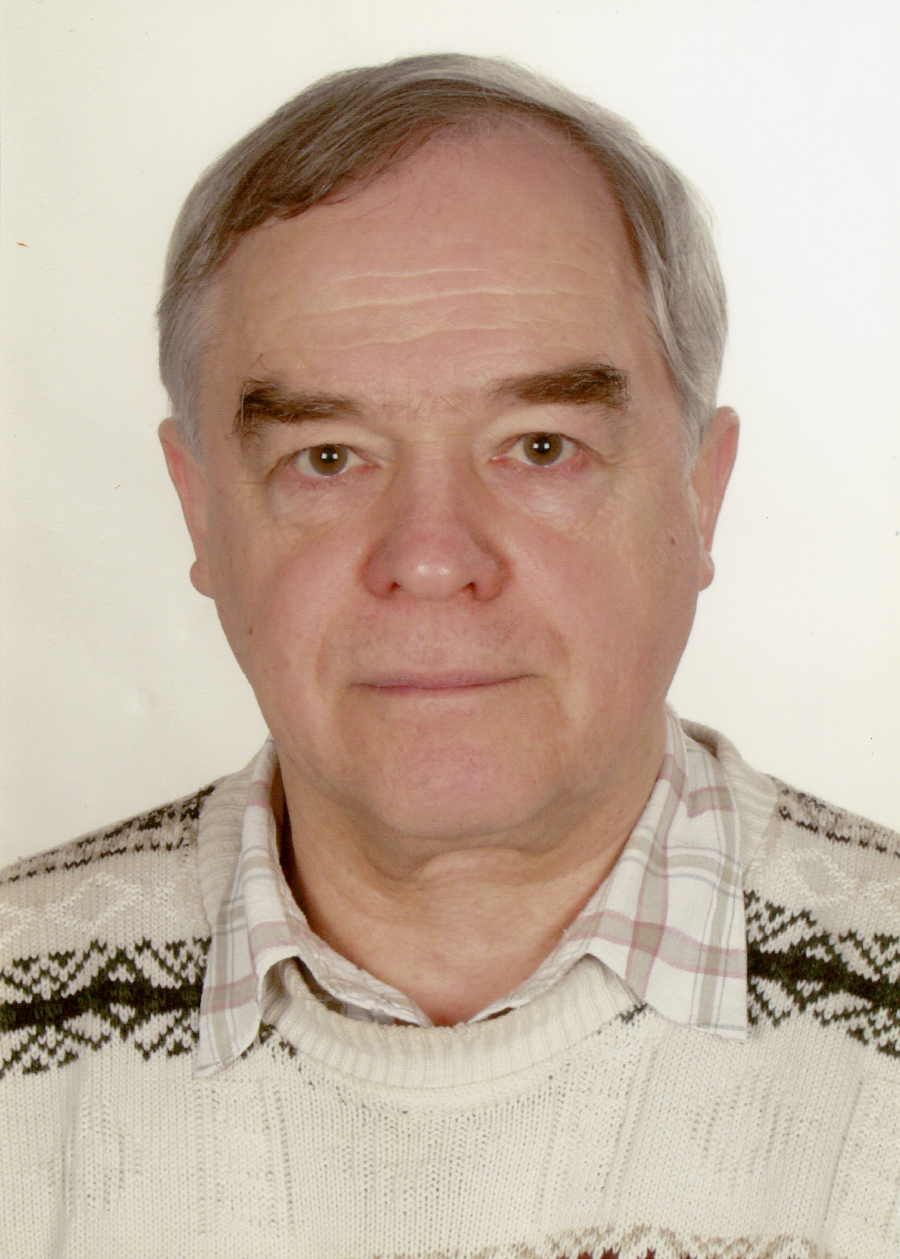}}]{Vyacheslav I. Yukalov}

graduated from the Physics Faculty of the Moscow State
University in 1970, with an M.Sc. in Theoretical Physics. He received his Ph.D.
in Theoretical and Mathematical Physics from the same University in 1974. He
also received the degree of Dr.Hab. in Theoretical Physics from the University
of Poznan and the degree of Dr.Sci. in Physics and Mathematics from the
Higher Attestation Committee of Russia.

Starting his career as a Graduate Assistant at the Moscow State University
(1970--1973), he then worked as Assistant Professor, Senior Lecturer, and
Associate Professor at the National Nuclear University of Moscow (1973--1984).
Since 1984 he has worked at the Joint Institute for Nuclear Research in Dubna
as Senior Scientist, Department Head, and where he currently holds the position
of Leading Scientist.

His research interests involve decision theory, quantum theory, dynamical theory,
complex systems, nonlinear and coherent phenomena, self-organization, and
development of mathematical methods.

His scientific awards include: Research Fellowship of the British Council,
Great Britain (1980--1981); Senior Fellowship of the University of Western
Ontario, Canada (1988); First Prize of the Joint Institute for Nuclear Research,
Russia (2000); Science Prize of the Academic Publishing Company, Russia (2002);
Senior Fellowship of the German Academic Exchange Program, Germany (2003); and
Mercator Professorship of the German Research Foundation, Germany (2004--2005).

V.I. Yukalov is the author of 400 papers in refereed journals and of four
books. He is an editor of five books and special issues. He is a member of the
American Physical Society, American Mathematical Society, European Physical
Society, International Association of Mathematical Physics, and Oxford University
Society.
\end{IEEEbiography}

\begin{IEEEbiography}[{\includegraphics[width=1in,height=1.25in,clip,keepaspectratio]{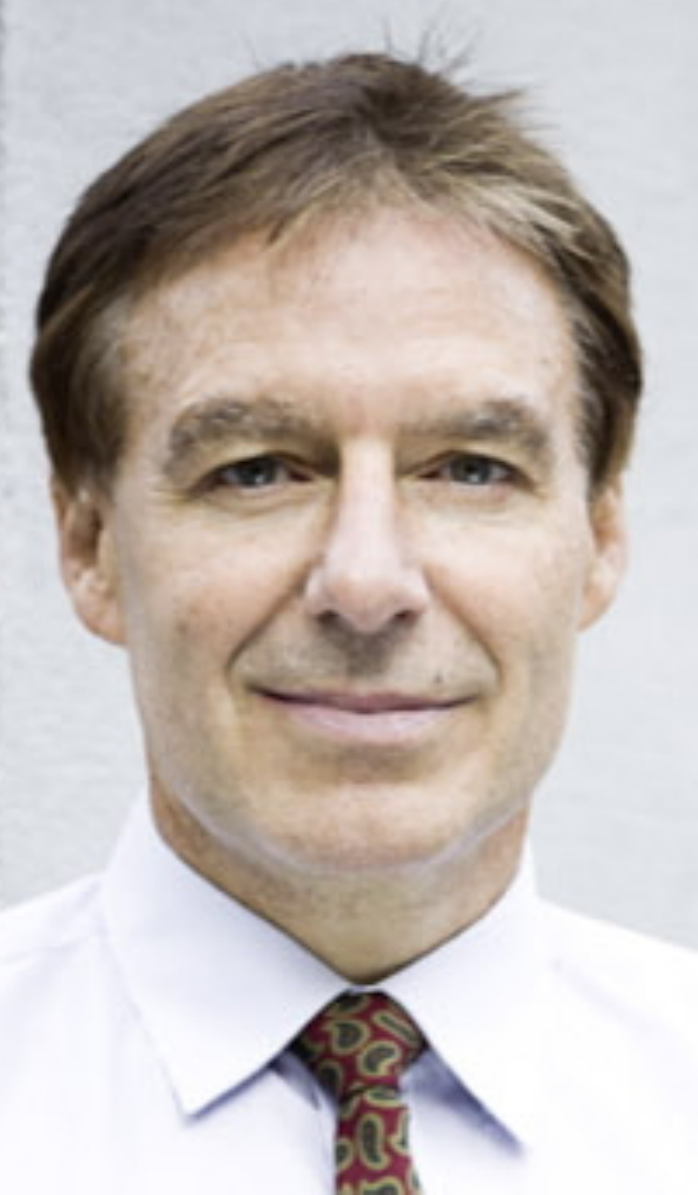}}]{Didier Sornette}

graduated from the Ecole Normale Sup\'erieure (ENS Ulm, Paris, France), in Physical Sciences in 1981, and obtained his PhD at the University of Nice, France in 1985.
His post-doctoral year in 1986 was in Paris, France, with Prof. P.-G. de Gennes on
complex matter and Prof. B. Souillard on Anderson localization.

He was Research Scientist at the french CNRS (National Center for Scientific Research)
from 1981 to 1990, a visiting scientist at Canberra, Australia (1984) and at the
Institute of Theoretical Physics of the University of Santa-Barbara in 1992, a
Research Director at CNRS from 1990 to 2006, a part-time Professor-in-Residence at
the Department of Earth and Space Sciences and at the Institute of Geophysics and
Planetary Physics, UCLA from Jan. 1996 to June 1999, and a Professor at UCLA from
July 1999 to February 2006. He was a Concurrent Professor of East China University
of Science and Technology (ECUST), Shanghai, China from May 2004 to March 2009, and
is now a Honorary Professor of the East China University of Science and Technology, Shanghai, China since 2009. Since March 2006, he is the Professor on the chair of
Entrepreneurial Risks at ETH Zurich (the Swiss Federal Institute of technology in
Zurich). He is also associated with the Department of Physics (D-PHYS) at ETH Zurich
since 2007 and with the Department of Earth Sciences (D-ERWD) at ETH Zurich since 2007.
He is also a professor of finance at the Swiss Finance Institute. He is the author
500+ research papers and 7 books. He uses rigorous data-driven mathematical statistical analysis combined with nonlinear multi-variable dynamical models including
positive and negative feedbacks to study the predictability and control of crises
and extreme events in complex systems, with applications to financial bubbles and
crashes, earthquake physics and geophysics, the dynamics of success on social
networks and the complex system approach to medicine (immune system, epilepsy and
so on) towards the diagnostic of systemic instabilities. In 2008, he launched the
Financial Crisis Observatory to test the hypothesis that financial bubbles can be
diagnosed in real-time and their termination can be predicted probabilistically.

Prof. Dr. D. Sornette was a recipient of the Science et D\'efence French Young
Investigator National Award (1985), the 2000 Research McDonnell award, the
Risques-Les Echos prize in 2002. He gave the E.N. Lorenz Lecture of the American Geophysical Union (AGU), December 16, 2010, the Ehrenfest Colloquium, Leiden, The Netherlands, 12 October 2011, and has been elected as a fellow of the AAAS (American Association for the Advancement of Science) since October 2013.

\end{IEEEbiography}

\end{document}